\documentclass[12pt,preprint]{aastex}

\newdimen\minuswidth    
\setbox0=\hbox{$-$}
\minuswidth=\wd0
\catcode`@=\active
\def@{\kern\minuswidth}
\newdimen\digitwidth    
\setbox0=\hbox{\rm0}
\digitwidth=\wd0
\catcode`!=\active
\def!{\kern\digitwidth}

\begin{document}

\shorttitle{Empirical metallicity, reddening and distance for old 
stellar populations}   
\shortauthors{Ferraro, Valenti \& Origlia}

\title{An empirical tool to derive
metallicity, reddening and distance for old stellar populations from 
near IR Color Magnitude Diagrams   
\footnote{Based on observations collected at the European Southern Observatory
(ESO), La Silla, Chile. Also based on observations made with
the Italian Telescopio Nazionale Galileo (TNG) operated on the island La Palma
by the Fundacion Galileo Galilei of INAF (Istituto Nazionale di Astrofisica) at
the Spanish Observatorio del Roque de los Muchachos of the Instituto de
Astrofisica de Canarias.}
}  

\author{Francesco R. Ferraro\altaffilmark{2},
Elena Valenti\altaffilmark{2,3},
 Livia Origlia\altaffilmark{3}}

\affil{\altaffilmark{2} Dipartimento di Astronomia Universit\`a di
Bologna, via Ranzani 1, I--40127 Bologna, Italy; francesco.ferraro3@unibo.it; 
elena.valenti3@unibo.it}  
\affil{\altaffilmark{3}
INAF--Osservatorio Astronomico di Bologna, via Ranzani 1, I--40127
Bologna, Italy; livia.origlia@bo.astro.it}

\begin{abstract}   
We present an empirical method to derive photometric metallicity,
reddening and distance to
old stellar populations by using a few major 
features of the Red Giant Branch (RGB) in near IR color magnitude diagrams. 
We combine the observed RGB features with  
a set of equations linking the global metallicity [M/H] 
to suitable RGB parameters (colors, magnitudes and slope),
as calibrated from a homogeneous sample of Galactic Globular
Clusters with different metallicities.
This technique can be applied to efficiently
derive the main population parameters of old stellar systems, 
in the view of using ground-based adaptive optics and space facilities 
to probe the stellar content of remote galaxies.

\end{abstract}  


\section{Introduction}   
\label{sec:intro}  

Stellar populations are fundamental tracers of the formation 
and evolution history of their parent galaxy.
The accurate determination of their basic parameters (as
age, metallicity and distance) are then crucial,
but this is restricted to a number of
relatively nearby stellar systems, for which single stars
are spatially resolved and whose magnitudes can be
accurately determined. At least the brightest portion of the 
Red Giant Branch (RGB) is needed in order to have hits on
metallicity and distance, while the detection of the Main
Sequence Turn-Off (MS\--TO) is required to get the age.
In this respect, detailed investigations of simple stellar
populations (SSP, {\it i.e.} coeval and chemically
homogeneous stellar aggregates) 
offer a unique opportunity to empirically calibrate 
suitable photometric indices and evolutionary features in
terms of the overall metallicity of the system. 

Stellar clusters represent the best approximation 
of SSP known in the Universe, hence they are ideal tools 
for this purpose.
An empirical method 
to get simultaneously metallicity and
reddening from the morphology and location of the RGB in
the $\rm (I,V-I)$ color-magnitude diagram (CMD) 
has been presented a decade ago \citep{SAR94}.
More recently, this method has been extended to the $\rm (V, B-V)$ plane by
\citet{sarlay97} and further improved by 
\citet{CB98} and \citet{F99} by adopting the \citet{cg97}
 and the global ($\rm [M/H]$) metallicity scale, respectively.
The method presented here adopts the most recent calibrations  
as obtained from a large IR photometric 
database of GCs, collected by our group  with different
instruments at the ESO telescopes (La Silla, Chile) and at the 
TNG (La Palma, Spain) over the last decade 
in the framework of a long-term project devoted to study the
photometric properties of the RGB \citep{F94,F95,M95,F00,S04,V04,Va04,Vb04}. 
In particular, we have recently presented a complete set of 
photometric indices (colors, magnitudes and slopes) describing
the location  and the morphology of the RGB and their
calibrations in terms of the global cluster metallicity
\citep{Va04}, together with the empirical calibrations of the 
IR luminosity of the two major RGB evolutionary features,
namely the RGB Bump and Tip \citep{Vb04}. It is worth mentioning that,
the calibration relations used in this study rely on a database,
whose properties have been derived in a fully self--consistent way. In fact,
the adopted estimates of the cluster reddening, metallicity
 and distance are based {\it i)} on a homogeneous photometric system; 
{\it ii)} on the \citet{F99} distance scale which relies
on the most recent and largest database of Galactic GCs; and 
{\it iii)} on a uniform and high--resolution metallicity scale \citep{cg97}.

The calibration of suitable relations to derive metallicity, reddening 
and distance in the near IR plane is crucial in the study of extragalactic 
bulges, which can be characterized by high metallicities and can be affected 
by severe reddening.
The current generation of ground-based IR instrumentation 
with high resolution and wide field coverage,  
and the future availability of the James Webb
Space Telescope will allow us to resolve the brightest
giants in galaxies up to several Mpc away, and to
derive their overall metallicity, reddening and distance 
modulus with great accuracy. 

The paper is organized as follows. \S 2 discusses the adopted metallicity
scale, while \S 3 and \S 4 present the equation sets for both
the {\it disk\--like} and {\it bulge\--like enrichment scenarios}, respectively. \S 5
describes the computational routine which provides  
photometric estimates of metallicity, reddening and distance, while in \S 6 we
test the described method on two template GCs in the Galactic 
bulge and in the  Large Magellanic 
Cloud, namely NGC~6539
and NGC~1978 respectively, in order to demonstrate its
reliability.

\section{The global metallicity}

As widely discussed in \citet{F99,F00} a correct
parameterization of the RGB characteristics as 
function of the metal content of the population does
require the knowledge of the so-called {\it global}
metallicity, which takes into account the iron
as well as the $\alpha$--element abundances. 
This is an important point since the
location of the RGB strongly depends on the low--ionization
potential [Fe+Mg+Si/H] abundance mixture \citep{SC91,SC96}
rather than on [Fe/H] abundance alone. In fact, 
low--ionization potential elements are
the main contributors to free electrons  which generate the
$\rm H^-$ ion, the major component
responsible for the continuum opacity in the
RGB temperature range 
\citep[3000--6000~K,][]{R77}.

In halo/disk field stars 
the average [$\alpha$/Fe] abundance ratio
shows a general enhancement of 0.3--0.5 dex with respect to the Solar
value up to [Fe/H]$\approx$--1 
\citep[see e.g.][ and references therein]{BOE99,gra00,car00}
and a linear decreasing trend towards
Solar [$\alpha$/Fe] with further increasing metallicity.
A [$\alpha$/Fe] enhancement is also found in the metal poor halo GCs 
\citep[see e.g.][ and references therein]{gra04,sne04}.
However, until a few years ago, only a few measurements were available 
in the high metallicity regime, to properly define the [$\alpha$/Fe] trend in GCs
\citep[see e.g.][]{kra94,car96}.
The actual position of the knee (i.e. the
metallicity at which [$\alpha$/Fe] begins to decrease) depends on the
type~Ia SN timescales and it is also a function
of the star formation rate, while the amount of
$\alpha $ enhancement depends on the initial mass function of 
the progenitors of the type~II SNe \citep[see][]{mw97}.

Measurements of metal rich field and cluster giants towards the Galactic bulge 
are a recent accomplishment
of high resolution optical and IR spectroscopy.
Bulge stars are indeed ideal targets to study the behavior
of the abundance patterns in the high metallicity domain,
but foreground extinction is so great
as to largely preclude optical studies of any kind, particularly
at high spectral resolution.
The most accurate abundance determinations
obtained so far and based on
high resolution optical spectroscopy refer
to a sample of K--giants in the Baade's window
\citep{MWR94,RMW00,ful04}
and a few giants in two GCs, namely NGC~6553 and NGC~6528
\citep{car01,zoc04}.
Recently, high resolution IR spectroscopic measurements of M giants in the 
Baade's window \citep{ro05} as well as of bulge 
GCs \citep{ori02,mel03,ori04,lcb04,orig05,ori05}
dramatically extended the sample of measured bulge stars.
All these studies point toward a $\alpha$-enhancement by a factor of 2-3 
up to solar metallicity.

\section{The  adopted enrichment scenarios
\label{scena}}

\begin{figure}[t!] 
\plotone{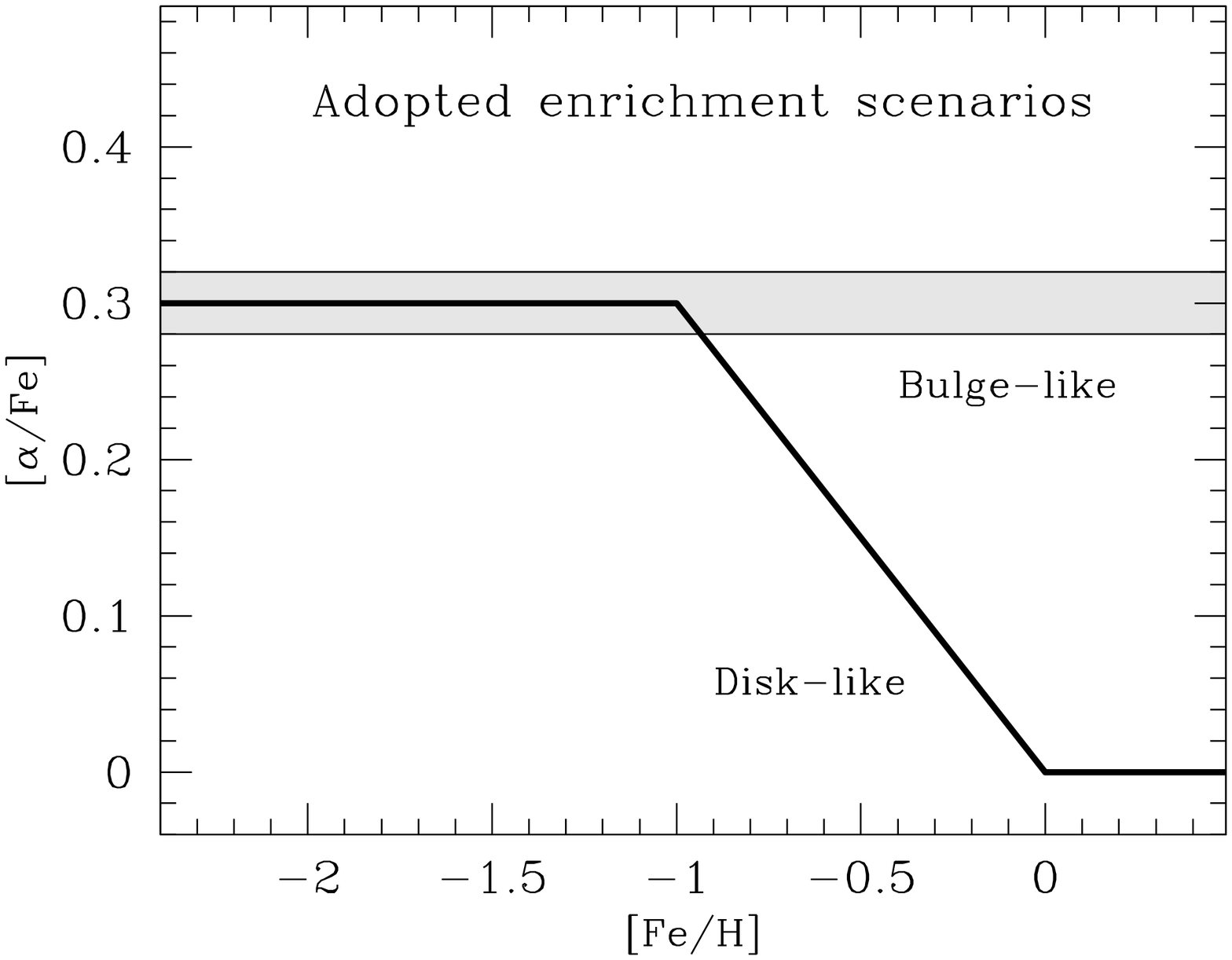}   
\caption{Sketched diagram of the [$\alpha$/Fe] {\it vs} [Fe/H] trend in 
the disk/bulge -like enrichment scenarios, as adopted in our computational 
routine.}
\label{scenario}  
\end{figure}

In the computation of the {\it global} metallicity for a sample of 61 GGCs, 
\citet{F99} used a constant
$\rm [\alpha/Fe]=0.28$ for $\rm [Fe/H]<-1$ and linearly decreasing
to zero at $\rm -1<[Fe/H]<0$.
In the following we compute two independent sets of global metallicities and  
photometric relations, according to two different scenarios 
(see Fig.~\ref{scenario}): 

{\it Scenario (1) -}  {\it Disk-like} enhancement 
(accordingly to \citet{F99}) 
$\rm [\alpha/Fe]=0.3$ for $\rm [Fe/H]<-1$ and $\rm [\alpha/Fe]$
linearly decreasing to zero for $\rm -1<[Fe/H]\le0$.

{\it Scenario (2) -}  {\it Bulge-like} enhancement  
(accordingly to Carney's (1996) suggestions and the recent results 
on the bulge field and cluster populations)
$\rm [\alpha/Fe]=0.30$ constant over the entire range of
metallicity ($\rm -2\le[Fe/H]\le0$).

The set of equations discussed in \citet[]{Va04}[hereafter Paper~I]
 and \citet[]{Vb04}[hereafter Paper~II]
have been computed accordingly to {\it Scenario (1)}. Here we 
present similar equations for  
{\it Scenario (2)}, so the reader can choose the 
equation set which turns out to be the most suitable to
the describe the stellar system.

In both scenarios the contribution of the
$\alpha$--element enhancement  has been taken into account
by simply re--scaling standard models to the {\it global}
metallicity [M/H] according to the following relation \citep{SCS93}:
\begin{equation}
\eqnum{2.1}
\label{met}
\rm [M/H] = [Fe/H] + log_{10} (0.638 f_{\alpha}+0.362)  
\end{equation}
where $f_{\alpha}$ is the enhanced factor of the
$\alpha$--elements.

\subsection{The equation set in the disk-like enrichment scenario
\label{diskeq}}

In this section we present and discuss the equation set adopted to construct the
computational routine which provides metallicity,
reddening  and distance of a stellar population, by using the
RGB observables in the $\rm (M_K,J-K)$ IR plane. 
Note that the validity of the relations has been extended up to
[M/H]${\sim}$+0.4~dex, with respect to the results presented in Paper~I,
including the recent results on the metallicity of NGC~6791 \citep{ori06}.
Hence
the relations presented here have been obtained in the metallicity range:
-2.16${\leq}$[Fe/H]${\leq}$+0.4 dex\footnote{Note that the inclusion 
of NGC6791 in our calibrator clusters sample
does not change significantly the relation derived in Paper I.}.

The overall metallicity of a stellar population can be linked to 
the slope of the RGB ($\rm slope_{RGB}$) as defined in
the near IR-CMD, by \citet{K95} and \citet{KF95}, 
accordingly to the following relation:
\begin{equation}
\eqnum{3.1}
\label{kslope_d}
\rm [M/H] = -2.53 -20.83 (slope^{JK}_{RGB})
\end{equation}

where $\rm slope^{JK}_{RGB}$ is
the RGB slope measured following the prescriptions
discussed in Paper~I in the  $\rm (M_K,J-K)$-CMD. 

The RGB\--tip luminosity is a quantity well
predicted by theoretical models and it can be easily
measured in populous stellar systems like galaxies.
The luminosity of the RGB tip for stellar populations
older than $t \sim 1-2$ Gyr is nearly  independent
from the population age. 
Moreover, it turns out to be particularly bright in the near\--IR 
($\rm M_K \sim -6$) hence, it is a very promising    
standard candle up to large distances.
The main limitation of this method is the
clear\--cut determination  of the RGB\--tip luminosity
that could be seriously affected by low\--number statistics. 
In fact, the rapid star evolution near the end of the RGB,
makes the brightest portion of the RGB intrinsically poorly
populated. However, this problem is marginal  
in very populous stellar systems.
The dependence of this feature from the overall metallicity
of the stellar population
has been empirically determined by using a sample of 
Galactic GCs (see Paper~II) and it 
turns out to be in good
agreement with theoretical expectations. Here we report the
relation obtained in Paper~II:
\begin{equation}
\eqnum{3.2}
\label{ktip_d}
\rm  M^{Tip}_K = -6.92 -0.62 [M/H] 
\end{equation}

As well known \citep[see][]{salgir02,grosara02}
for clusters older than 2 Gyr the level of the 
Helium-burning Red clump (RC) is mainly influenced
by metallicity and it shows little dependence on the age. Hence
for relatively old  metal rich populations ($\rm t>2~Gyr, [M/H]>-1$) 
it is also possible to use the level of the RC as 
an additional distance indicator. 
We estimate the mean level of the RC by using a metal-rich
(-0.9$<$[Fe/H]$<$-0.3) cluster subsample (namely, 47~Tuc, NGC~6342, NGC~6380,
NGC~6441, NGC~6440, NGC~6553, NGC~6528, and NGC~6637) selected from the global database
presented in PaperI. The value turns out to be:  
\begin{equation}
\eqnum{3.3}
\rm M_K^{RC}= -1.40\pm0.2
\end{equation}
Note that most selected clusters can be considered  coeval within 10-20\% 
accordingly to \citet{deang05} who measured the relative ages of
47~Tuc, NGC~6342, NGC~6637 and \citet{ort95,ort01} who measured
   the relative ages of 47~Tuc, NGC~6528, NGC~6553.
 The adopted uncertainty of $\pm 0.2$ in equation (3), takes  into account the  
 metallicity dependence of the RC position in the metallicity range 
 spanned by our metal--rich GC subsample ($\rm \delta[Fe/H]\sim 0.6 dex$). 
 In fact, accordingly to Table 1  by \citet{salgir02} the  
$\rm M_K^{RC}$ level is expected to vary $\rm \sim 0.15 mag$ at $\rm t=10 Gyr$
 over a such metallicty range.  Moreover in the age range 2-10Gyr, 
 at fixed metallicity ($\rm [M/H]=-0.68$) the $\rm M_K^{RC}$ 
 varies 0.03 mag per Gyr.
 Hence the 0.2 mag uncertainty adopted as conservative estimate of the 
 $\rm M_K^{RC}$ level
is expected to be reasonable over the considered range of age and metallicity.

The absolute magnitude of the RGB at fixed color
is another observable that can be used to
define useful relations with the overall metallicity.
Here we report the calibrations discussed in Paper~I:
\begin{equation}
\eqnum{3.4}
\label{kmag_d}
\rm M^{(J-K)_0=0.7}_K = -1.38 +2.22 [M/H] 
\end{equation}

Finally, the entire set of equations describing the
RGB location in $\rm (J-K)$ color at 
different level of magnitudes 
($\rm M_K=-5.5,-5,-4,-3$) are listed.
\begin{equation}
\eqnum{3.5}
\label{kcol55_d}
\rm [M/H] = -4.37+3.84(J-K)^{-5.5}_0 
\end{equation}
\begin{equation}
\eqnum{3.6}
\label{kcol5_d}
\rm [M/H] = -4.51+4.24(J-K)^{-5}_0
\end{equation}
\begin{equation}
\eqnum{3.7}
\label{kcol4_d}
\rm [M/H] = -4.87+5.20(J-K)^{-4}_0
\end{equation}
\begin{equation}
\eqnum{3.8}
\label{kcol3_d}
\rm [M/H] = -5.36+6.48(J-K)^{-3}_0
\end{equation}

\subsection{The equation set in the bulge-like enrichment scenario
\label{bulge}}

An analogous set of 16 equations can be obtained in the
{\it bulge-like} enrichment scenario. Here we list the complete equation set
based on the database presented in Paper~I and II by adopting a constant
$\rm [{\alpha}/Fe]$ over the entire range of metallicity.
\begin{equation}
\eqnum{4.1}
\label{kslope_b}
\rm [M/H] = -2.63 -22.50 (slope^{JK}_{RGB})
\end{equation}
\begin{equation}
\eqnum{4.2}
\label{ktip_b}
\rm M^{Tip}_K= -6.84 -0.56 [M/H]
\end{equation}
\begin{equation}
\eqnum{4.3}
\label{khb_b}
\rm  M^{RC}_K = -1.40\pm 0.2
\end{equation}
\begin{equation}
\eqnum{4.4}
\label{kmag_b}
\rm M^{(J-K)_0=0.7}_K = -1.58 +2.08 [M/H] 
\end{equation}
\begin{equation}
\eqnum{4.5}
\label{kcol55_b}
\rm  [M/H] = -4.59 +4.13(J-K)^{-5.5}_0 
\end{equation}
\begin{equation}
\eqnum{4.6}
\label{kcol5_b}
\rm  [M/H] = -4.74 +4.55(J-K)^{-5}_0
\end{equation}
\begin{equation}
\eqnum{4.7}
\label{kcol4_b}
\rm  [M/H] = -5.12 +5.58(J-K)^{-4}_0
\end{equation}
\begin{equation}
\eqnum{4.8}
\label{kcol3_b}
\rm  [M/H] = -5.64 +6.94(J-K)^{-3}_0
\end{equation}

\section{The computational routine}  

\begin{figure}[t!]
\plotone{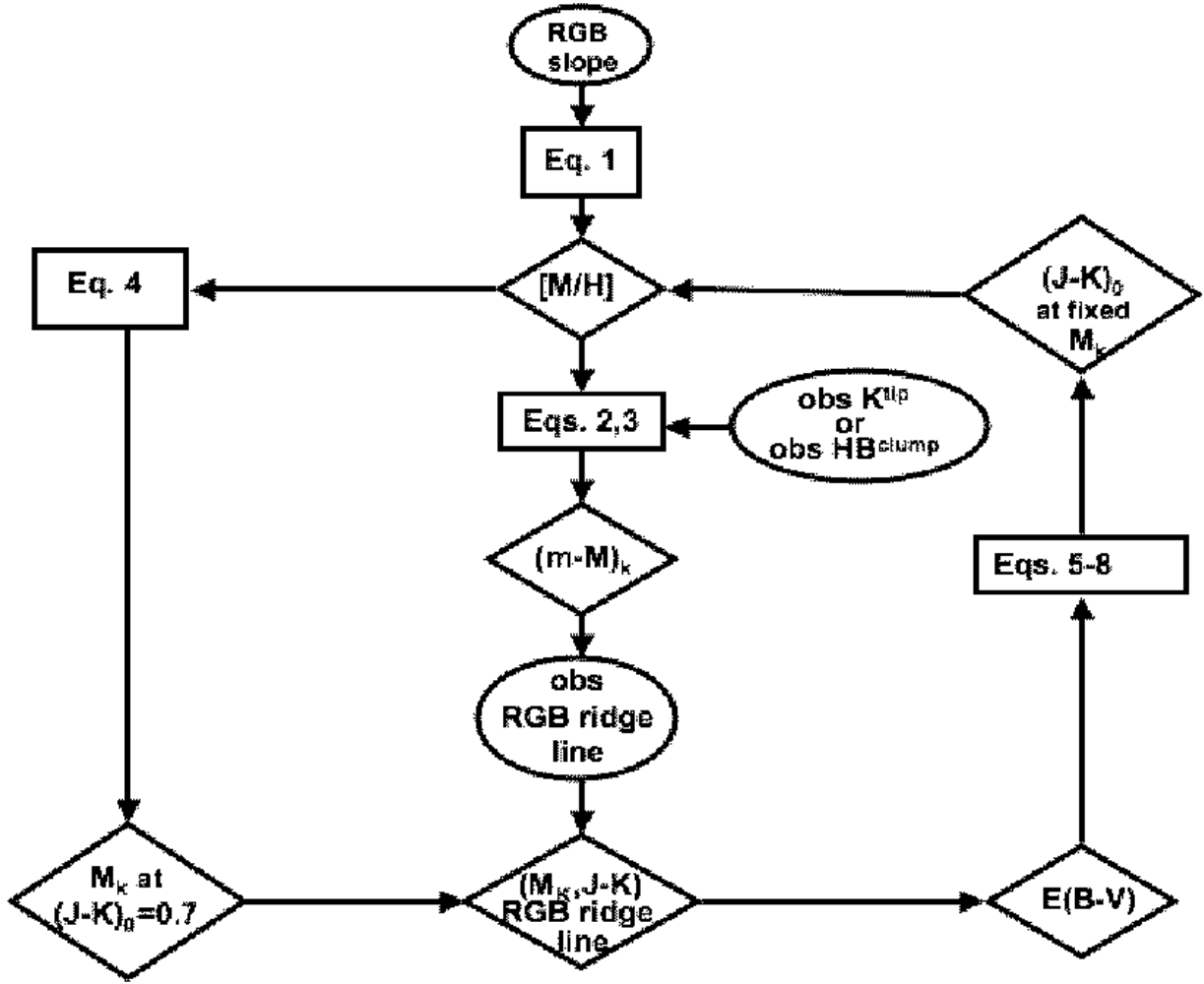}   
\caption{Flow diagram of the computational routine used to 
derive simultaneously distance, reddening and metallicity of an 
old stellar population from the RGB morphology (slope, colors, luminosity) 
and evolutionary features (RGB-Tip and  red clump) in the near IR CMDs.}
\label{diagram}  
\end{figure}

The computational routine requires as input parameters: 
\begin{itemize}
\item the observed RGB mean ridge line 
in the $\rm (K,J-K)$ CMD; 
\item the observed RGB slope 
($\rm slope^{JK}_{RGB}$); 
\item the observed RGB-Tip
($\rm K^{obs}_{Tip}$) and/or the RC level
($\rm K^{obs}_{RC}$). 
\end{itemize}
Fig.~\ref{diagram} shows the flow-diagram of the computational routine.
In the following, references to eqs.~1\--8  either
equations in Sect.~\ref{diskeq} or in Sect.~\ref{bulge}, depending 
on the adopted enrichment scenario.

Since the RGB slope is independent
of the cluster distance and reddening, we can use eq.~(1)
to get a first guess of the population metallicity.  
By using this first guess metallicity, it is
possible to derive the expected level of the RGB Tip from
eq.~(2). Hence,  the comparison with the observed
value ($\rm K^{obs}_{Tip}$) will directly yield
the first estimate of the distance modulus $\rm (m-M)_K$.
For the (closest) metal-rich  stellar systems
for which the RC is observed, an independent estimate of
the modulus can be also derived from the comparison with  
the mean RC levels reported in eq.~(3).

By using  the first guess metallicity
and eq.~(4) it is also possible to obtain
the expected absolute magnitude  ($\rm M_K^{exp}$) 
at fixed color ($\rm (J-K)_0=0.7$). 

Now, the mean ridge line
of the program population can be corrected by
using the distance modulus obtained above: the observed
color ($\rm (J-K)^{obs}$) in correspondence of the derived
magnitude ($\rm M_K^{exp}$) will yield the first
guess estimate of the reddening:
$$\rm  E(B-V)= ((J-K)^{obs}-0.7)/(0.87-0.38)$$
where the extinction coefficients from \citet{SM79}
have been adopted.

The color of the input RGB mean ridge line can be now
de\--reddened and transformed into the absolute plane 
$\rm (M_K,(J-K)_0)$.  
Once in the absolute plane, 
the color at fixed magnitude levels can be measured and 
inserted in eqs.~(5\--8) to get a new mean value of 
metallicity.
The latter can be inserted in eq.~(2), 
and the overall procedure iterated 
until the values of the three output quantities, namely metallicity,
reddening and distance, converge within suitable ranges, 
which can be specified as input tolerances.

The formal error on each derived quantity can be estimate by using a 
simple Monte Carlo simulation. In doing this we randomly extracted 
a large number of values for each of the
three main observables (namely  $\rm slope^{JK}_{RGB}$, $\rm K^{Tip}_{obs}$ and 
$\rm K^{obs}_{RC}$) from a Gaussian distribution peaked on the 
observed value  and with
$\sigma$ equal to the  uncertainty of each observable. 
These values are used  (following
the scheme plotted in Fig.~\ref{diagram}) to derive  metallicity, 
reddening and  distance. The
standard deviation of each set of values is the error associated 
to that specific quantity. However, beside the formal errors obtained 
from this procedure
(typically $\rm \delta(m-M)_0=\pm0.10-0.15$ mag;  
$\rm \delta E(B-V)=\pm0.03-0.04$ mag; $\rm \delta[M/H]=\pm0.10-0.15$ dex)
we estimate that conservative uncertainties for the derived quantities are:
$\rm \delta(m-M)_0=\pm0.2$ mag; $\rm \delta E(B-V)=\pm0.05$ mag; 
$\rm \delta[M/H]=\pm0.2$ dex.

\section{Test cases: NGC~1978 and NGC~6539}
\begin{figure}[t!]    
\plotone{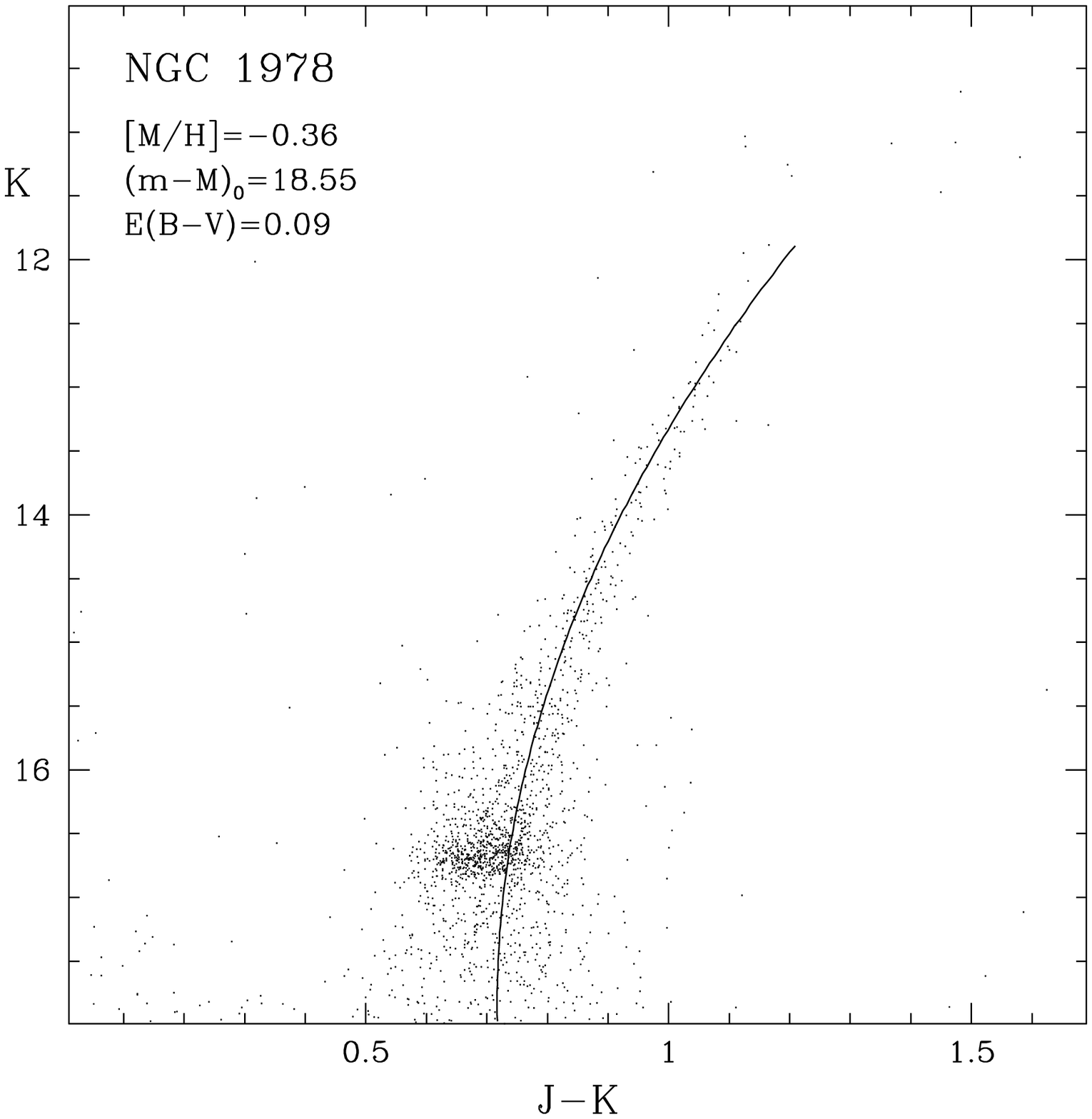}
\caption{Observed K, (J\--K) Color\--Magnitude Diagram of NGC~1978. 
The derived RGB fiducial ridge line is overplotted as solid lines.
The derived metallicity, distance and reddening solutions within the {\it
disk\--like scenario} are also reported.}
\label{1978}
\end{figure}

\begin{figure}[t!]    
\plotone{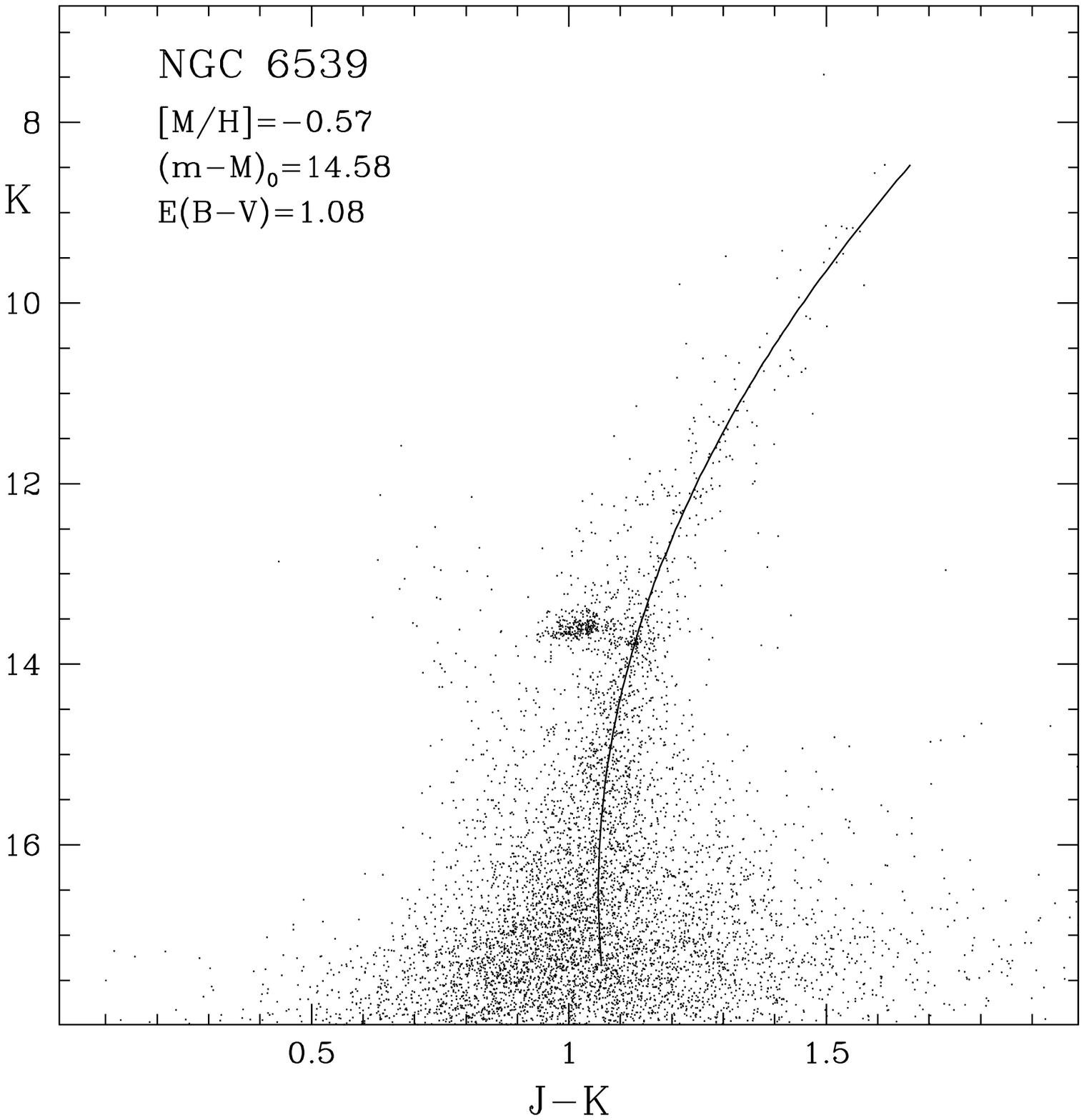}
\caption{Observed K, (J\--K) Color\--Magnitude Diagram of NGC~6539. 
The derived RGB fiducial ridge
lines are overplotted as solid lines.
The derived metallicity, distance and reddening solutions within the {\it
bulge\--like scenario} are also reported.}
\label{6539}
\end{figure}

To test the reliability of the proposed technique we have chosen two
metal\--rich clusters for which accurate determination of the metallicity have
been recently obtained: the Large Magellanic Cloud cluster NGC~1978, and the
Galactic bulge cluster NGC~6539. The CMDs of NGC~1978 and NGC~6539 shown in
Fig.~\ref{1978} and ~\ref{6539} are from \citet{muc06} and from
\citet{ori05}, respectively.
The RGB mean ridge lines computed following the prescriptions described in
Paper~I are also overplotted as solid line onto the CMDs of 
Fig.~\ref{1978} and ~\ref{6539}.

In the case of NGC~1978, the CMD shown in Fig.~\ref{1978} has been used to
measure the RGB slope 
($\rm slope^{JK}_{RGB}$=--0.104) and to 
estimate the observed RGB\--Tip (K$\rm ^{obs}_{Tip}$=11.89). 
It is worth noticing
that this cluster is suspected to be a relatively young cluster
 ($\rm t\sim 2 Gyr$), hence  we do not use its RC level 
 in deriving the distance modulus\footnote{Only the $\rm K^{Tip}_{obs}$ 
 level has been used for this cluster.}
since for extreme young clusters the  $\rm
  K^{obs}_{RC}$ level is a sensible function  of the age 
of the population \citep[see e.g. Fig.~3 and Table~1 by][]{salgir02}. 

By running the computational routine with the input parameters quoted above, a
good convergence is reached after few iterations.
In the {\it bulge\--like scenario}, the computational routine leaded a reddening
estimate of E(B\--V)=0.09$\pm$0.04~mag, an intrinsic distance modulus of
(m-M)$_0$=18.53$\pm$0.18~mag and a metallicity of [M/H]=--0.29$\pm$0.14~dex. 
In the {\it
disk\--like scenario} the following solutions have been obtained:  
E(B\--V)=0.09$\pm$0.04~mag, (m\--M)$_0$=18.55$\pm$0.19~mag, 
and [M/H]=--0.36$\pm$0.12~dex.
These values are fully
consistent with the results found by \citet{per83} and \citet{sch98} 
for the extinction
(E(B\--V)=0.100, and E(B\--V)=0.075, respectively), by
\citet{vdb98} and \citet{alves} for the distance ((m-M)$_0$=18.48 and
(m-M)$_0$=18.50, respectively), and by \citet{frf06} and \citet{olsz} for the
metallicity ([Fe/H]=--0.38~dex and [Fe/H]=--0.42~dex, based on 
high\--resolution optical spectra and on the Calcium Triplet, respectively).

Adopting the same strategy followed in the case of NGC~1978, we 
have analyzed the CMD of NGC~6539 
\citep{ori05}, finding a slope$\rm ^{JK}_{RGB}$=--0.091
and K$\rm ^{obs}_{Tip}$=8.47~mag.
From the derived K luminosity function we also 
estimated the RC level to be K$\rm ^{obs}_{RC}$=13.60 mag. 
By running the computational routine 
within the {\it bulge\--like scenario},
we found a reddening of E(B\--V)=1.08$\pm$0.06~mag,
an intrinsic distance modulus of $(m\--M)_0=14.58\pm0.15$~mag, 
and a global metallicity
[M/H]=--0.57$\pm$0.15~dex. The derived metallicity is fully consistent within 
0.2~dex with the recent estimate, based on high\--resolution
(R${\approx}25000$) IR spectroscopy, by \citet{ori05}, who found
[Fe/H]=--0.76~dex. The obtained reddening value is also in excellent agreement
with the estimate listed in the \citet{harris} compilation (E(B\--V)=1.00), and
with the \citet{sch98} extinction map (E(B\--V)=1.09).
Our distance value turns out to be 0.1~mag larger than the
\citet{harris} estimate\footnote{Note that the adopted calibrations are 
based on the \citet{F99} distance scale, which is
$\approx$0.15\--0.20 mag longer that the one by \citet{harris}.}.

\acknowledgements  

This research 
is supported by the Agenzia Spaziale Italiana (ASI) and the 
Ministero dell'Istruzione, dell'Universit\`a e della Ricerca. 
We thank the anonymous referee for his/her helpful comments and suggestions which
significantly contributed to improve the paper presentation.

\clearpage 

\clearpage

\clearpage 

\clearpage 

\begin{thebibliography}{}  
\bibitem[Alves (2004)]{alves}
Alves, D.~R. 2004, NAR, 48, 659
\bibitem[Barbuy et al.(1997)]{barb97}
Barbuy, B., Ortolani, S. \& Bica, E. 1997, \aap S, 122, 483
\bibitem[Boesgaard et al.(1999)]{BOE99}
Boesgaard, A. M., King, J. R., Deliyannis, C. P., \& Vogt, S. S. 
1999, \aj, 117, 492
\bibitem[Carney(1996)]{car96}
Carney, B. W. 1996, \pasp, 108, 877
\bibitem[Carretta \& Gratton(1997)]{cg97}
Carretta, E. \& Gratton, R.~G. 1997, A\&AS, 121, 95
\bibitem[Carretta \& Bragaglia(1998)]{CB98}
Carretta, E., \& Bragaglia, A. 1998, 329, 937
\bibitem[Carretta, Gratton \& Sneden(2000)]{car00}
Carretta, E., Gratton, R. G., \& Sneden, C. 2000, \aap, 356, 238
\bibitem[Carretta et al.(2001)]{car01}
Carretta, E., Cohen, J., Gratton, R.G., \& Behr, B.  2001,
\aj,  122, 1469
\bibitem[De~Angeli et al.(2005)]{deang05}
De~Angeli, F. et al. 2005, \aj, 130, 116
\bibitem[Fahlman et al.(1997)]{fah95}
Fahlman, G.~G., Douglas, K.~A, Thompson, I.~B., 1995, \aj, 110, 2189
\bibitem[Ferraro et al.(1994)]{F94} 
Ferraro, F. R., Fusi Pecci, F., Guarnieri, M. D., Moneti, A., Origlia, L., 
\&Testa, V. 1994, \mnras, 266, 829
\bibitem[Ferraro et al.(1995)]{F95} 
Ferraro, F. R., Fusi Pecci, F., Montegriffo, P., Origlia, L., 
\& Testa, V. 1994, \aap, 298,461
\bibitem[Ferraro et al.(1999)]{F99} 
Ferraro, F.R., Messineo, M., Fusi Pecci, F., De Palo, M.A., Straniero, O.,
Chieffi, A. \& Limongi, M., 1999, \aj, 118, 1738
\bibitem[Ferraro et al.(2000)]{F00} 
Ferraro F.R., Montegriffo P., Origlia L., Fusi Pecci F., 2000, \aj, 119, 1282
\bibitem[Ferraro et al.(2006)]{frf06}
Ferraro F.R., Mucciarelli, A., Carretta, E., Origlia, L. 2006, ApJ, {\it submitted}  
\bibitem[Fulbright, Rich \& McWilliam (2004)]{ful04}
Fulbright, J.P., Rich, R.M., \& McWilliam, A.
2004, AAS, 205, 7706
\bibitem[Lee, Carney \& Balachandran (2004)]{lcb04}
Lee, J., Carney, B. W., Balachandran, S. C. 2004,
\aj, 128, 2388
\bibitem[Gratton et al.(2000)]{gra00}
Gratton, R. G., Sneden, C., Carretta, E., \& Bragaglia, A. 2000, \aap, 354, 169
\bibitem[Gratton, Sneden \& Carretta(2004)]{gra04}
Gratton, R., Sneden, C., \& Carretta, E. 2004, \araa, 42, 385
\bibitem[Grocholski \& Sarajedini(2002)]{grosara02}
Grocholski, A.~J. \& Sarajedini, A. 2002, \aj, 123,1603
\bibitem[Harris(1996)]{harris}
Harris, W.~E. 1996, \aj, 112, 1487, for the 2003 updated version see
http://\-physwww.mcmaster.ca\-/\%7Eharris/\-mwgc.dat
\bibitem[Kraft(1994)]{kra94}
Kraft, R. P. 1994, \pasp, 106, 553
\bibitem[Kuchinski et al.(1995)]{K95}
Kuchinski, L. E., Frogel, J. A., Terndrup, D. M., \& Persson, S. E. 
1995, \aj, 109, 1131
\bibitem[Kuchinski \& Frogel(1995)]{KF95}
Kuchinski, L. E., \& Frogel, J. A. 1995, \aj, 110, 2844
\bibitem[McWilliam \& Rich(1994)]{MWR94}
McWilliam, A., \& Rich, R.M. 1994, \apjs, 91, 749
\bibitem[McWilliam(1997)]{mw97}
McWilliam, A. 1997, \araa, 35, 503
\bibitem[Mel\'endez et al.(2003)]{mel03}
Mel\'endez, J., Barbuy, B., Bica, E., Zoccali, M., Ortolani, S., Renzini, A.,
\& Hill, V. 2003, \aap, 411, 417
\bibitem[Montegriffo et al.(1995)]{M95}
Montegriffo, P., Ferraro, F. R., Fusi Pecci, F., \& Origlia, L. 
1995, \mnras, 276, 739
\bibitem[Mucciarelli et al.(2006)]{muc06}
Mucciarelli, A., Origlia, L., Ferraro, F.~R., Maraston, C., Testa, V.,
2006, \apj, in press, astro-ph/0604139
\bibitem[Olszewski et al.(1991)]{olsz}
Olszewski, E.~W., Schommer, R.~A., Suntzeff, N.~B., Harris, H.~C. 1991, \aj,
101, 515
\bibitem[Origlia, Rich \& Castro(2002)]{ori02}
Origlia, L., Rich, R. M., \& Castro, S. 2002, \aj, 123, 1559
\bibitem[Origlia \& Rich(2004)]{ori04}
Origlia, L., \& Rich, R. M. 2004, \aj, 127, 3422
\bibitem[Origlia , Valenti \& Rich(2005)]{orig05}
Origlia, L., Valenti, E. \& Rich, R. M. 2005, \mnras, 356, 1276 
\bibitem[Origlia et al. (2005)]{ori05}
Origlia, L., Valenti, E., Rich, R. M. \& Ferraro, F. R 2005, 
\mnras, 363, 897
\bibitem[Origlia et al. (2006)]{ori06}
Origlia, L., Valenti, E., Rich, R. M. \& Ferraro, F. R 2006, 
\mnras, in press, astro-ph/0604030
\bibitem[Ortolani et al.(1995)]{ort95}
Ortolani, S., Renzini, A., Gilmozzi, R., Marconi, G., Barbuy, B., Bica, E., 
Rich, R.~M. 1995, Natur, 337, 701
\bibitem[Ortolani et al.(2001)]{ort01}
Ortolani, S., Barbuy, B., Bica, E., Renzini, A., Zoccali, M., 
Rich, R.~M., Cassisi, S. 2001, \aap, 376, 878
\bibitem[Persson et al.(1983)]{per83}
Persson, S.~E., Aaronson, M., Cohen, J.~G., Frogel, J.~A., Matthews, K. 1983,
\apj, 266, 105
\bibitem[Renzini(1977)]{R77}
Renzini, A. 1977, in Advanced Stages in Stellar Evolution, 
ed. P. Bouvier \& A.  Maeder (Geneva: Geneva Obs.), 151
\bibitem[Rich \& Origlia(2005)]{ro05}
Rich, R.M., \& Origlia, L.  2005, \apj,634, 1293 
\bibitem[Rich \& McWilliam(2000)]{RMW00}
Rich, R.M., \& McWilliam, A.  2000, SPIE, 4005, 150
\bibitem[Salaris \& Cassisi(1996)]{SC96}    
Salaris, M., \& Cassisi, S. 1996, \aap, 305, 858
\bibitem[Salaris, Chieffi \& Straniero(1993)]{SCS93} 
Salaris, M., Chieffi, A., Straniero, O., 1993, \apj, 414, 580
\bibitem[Salaris \& Girardi(2002)]{salgir02}
Salaris, M. \& Girardi, L. 2002, \mnras, 337, 332
\bibitem[Sarajedini(1994)]{SAR94}
Sarajedini, A. 1994, \aj, 107, 618
\bibitem[Sarajedini \& Layden(1997)]{sarlay97}
Sarajedini, A. \& Layden, A. 1997, \aj, 113, 264
\bibitem[Savage \& Mathis (1979)]{SM79} 
Savage, B.D., Mathis, J.S., 1979, \araa, 17, 73
\bibitem[Schlegel et al.(1998)]{sch98}
Schlegel, D.~J., Finkbeiner, D.~P., Davis, M. 1998, \apj, 500, 525
\bibitem[Sneden et al.(2004)]{sne04}
Sneden, C., Kraft, R.P., Guhathakurta, P., Peterson, R.C.,Fulbright, J.P. 2004,
\aj, 127, 2162
\bibitem[Sollima et al.(2004)]{S04}
Sollima, A., Ferraro, F. R., Origlia, L., Pancino, E., \& Bellazzini, M. 
2004, \aap, 420, 173   
\bibitem[Straniero \& Chieffi(1991)]{SC91}
Straniero, O., \& Chieffi, A. 1991, \apjs, 76, 525
\bibitem[Valenti et al.(2004)]{V04}
Valenti, E., Ferraro, F. R., Perina, S., \& Origlia, L. 2004, \aap, 419, 139
\bibitem[Valenti, Ferraro \& Origlia(2004a)]{Va04}
Valenti, E., Ferraro, F. R., \& Origlia, L. 2004, \mnras, 351, 1204, {\it
Paper~I}
\bibitem[Valenti, Ferraro \& Origlia(2004b)]{Vb04}
Valenti, E., Ferraro, F. R., \& Origlia, L. 2004, \mnras, 404, 157, {\it
Paper~II}
\bibitem[Van Den Bergh (1998)]{vdb98}
Van Den Bergh, S. 1998, \pasp, 110, 1377
\bibitem[Zoccali et al.(2004)]{zoc04}
Zocali, M., Barbuy, B., Hill, V., Ortolani, S., Renzini, A.,
Bica, E., Momany, Y., Pasquini, L., Minniti, D., Rich, R. M. 2004,
\aap, 423, 507

\end{thebibliography}
\end{document}